\documentclass[12pt,a4paper]{article}

\usepackage{arxiv}

\usepackage[utf8]{inputenc}
\usepackage{pgfplots}
\usepackage{layouts}
\usepackage{amsmath}
\usepackage{array}
\usepackage{blindtext}
\usepackage{hyperref}
\usepackage{comment}
\usepackage{pgf}

\newcommand{\bfu}{\mbox{\boldmath $u$}}

\newcommand{\bfx}{\mbox{\boldmath $x$}}
\newcommand{\bfy}{\mbox{\boldmath $y$}}
\newcommand{\bfz}{\mbox{\boldmath $z$}}
\newcommand{\eps}{{\epsilon}}

\newcommand{\Gam}{{\Gamma}}

\newcommand{\Ome}{{\Omega}}
\newcommand{\bfeps}{\mbox{\boldmath $\epsilon$}}

\newcommand{\bfsig}{\mbox{\boldmath $\sigma$}}

\newcommand{\ca}{{\cal A}}

\newcommand{\cu}{{\cal U}}

\newcommand{\dist}{\mathrm{dist}}
\newcommand{\disth}{\mathrm{dist}^h}
\newcommand{\disthp}{\mathrm{dist}^{h+}}

\newcommand{\rmd}{\ensuremath{\mathrm{d}}}
\newcommand{\dint}{\mbox{\, \rmd}}

\newcommand{\normgrad}[1]{\| \nabla #1 \| }

\newcommand{\Yc}{Y_\mathrm{c}}

\newcommand{\Gc}{G_\mathrm{c}}

\newcommand{\dotd}{\dot{d}}

\newcommand{\texton}{\text{\ on\ }}

\newcommand{\abs}[1]{\lvert #1 \rvert}
\newcommand{\norm}[1]{\lVert #1 \rVert}

\newcommand{\dt}{\Delta t}
\newcommand{\du}{\overline{d}}
\newcommand{\dl}{\underline{d}}
\newcommand{\dloc}{d_\mathrm{loc}}
\newcommand{\od}{\overline{d}}
\newcommand{\oD}{\overline{D}}
\newcommand{\oca}{\overline{\ca}}

\newcommand{\Lip}{\mathcal{L}}
\newcommand{\Liph}{\mathcal{L}^h}
\newcommand{\Liphb}{\overline{\mathcal{L}}^h}
\newcommand{\Liphu}{\mathcal{L}^{h1}}
\newcommand{\Liphd}{\mathcal{L}^{h2}}
\newcommand{\Liphp}{\mathcal{L}^{h+}}
\newcommand{\Ltwo}{L^2(\Omega)}
\newcommand{\grad}{\nabla}
\newcommand{\Vt}{V_{\mathrm{trial}}}
\newcommand{\Vf}{V_{\mathrm{final}}}
\newcommand{\Omeh}{\Omega^h}

\title{Lipschitz regularization for fracture: \\ the Lip-field approach}
\author{N. Chevaugeon and N. Moës }

\author{Nicolas Chevaugeon \\
	Ecole Centrale de Nantes\\
	GeM Institute, UMR CNRS 6183 \\
	1 rue de la No\"{e}, 44321 Nantes, France \\
	\texttt{nicolas.chevaugeon@ec-nantes.fr} \\
	\And
	Nicolas Mo\"es \\
	Ecole Centrale de Nantes\\
	GeM Institute, UMR CNRS 6183 \\
	Institut Universitaire de France (IUF) \\
	1 rue de la No\"{e}, 44321 Nantes, France \\
	\texttt{nicolas.moes@ec-nantes.fr} 
	}

\begin{document}
\maketitle

\begin{abstract}
    The Lip-field approach is a new regularization method for softening material material models. It was presented first in \cite{Moes2021lipschitz} providing one-dimensional simulations for damage and plasticity. The present paper focuses on a two-dimensional implementation for elasto-damage models (quasi-brittle fracture). 
    The incremental potential used in the Lip-field approach is the non-regularized one. The regularization comes from the addition of a Lipschitz constraint on the damage field. 
    In other words, the free energy does not depend on the damage gradient.
    The search of the displacement and damage fields from one time-step to the next is based on an iterative staggered scheme. The displacement field is sought for a given damage field. Then, a Lipschitz continuous damage field is sought for a given displacement field. Both problems are convex. The solution to the latter benefits from bounds proven in \cite{Moes2021lipschitz} and used in this paper.
    The paper details the implementation of the Lipschitz regularity on a finite element mesh and details the overall solution scheme. Four numerical examples demonstrate the capability of the new approach.
\end{abstract}

\section{Introduction}

Fracture mechanics started with the work of A.A. Griffith \cite{Griffith2021}. The Griffith's crack model considers that a crack evolves as a point in 2D media or a curve in 3D media leaving behind crack faces across which displacements jumps are allowed. An energy is required for crack advance (toughness).  The Griffith model is not able to predict crack initiation  (an infinite load is predicted for a crack size going to zero) and is not able to predict branching cracks (one tip becoming two).
The Griffith model was later improved/generalized in at least two major ways: 
cohesive zone models and diffuse damage models.
Both models introduce a length scale in the fracture model which is absent in the Griffith model.

The cohesive zone model   (CZM)\cite{Dugdale1960,Barenblatt1961} recognizes that there exists a process zone ahead of the crack tip whose size may not always be neglected. For instance, 
it can up to 5 to 10 cm for concrete. The process zone introduces a length scale and size effect. The CZM is a quite popular model in computational mechanics. Cracks are allowed to initiate and propagate  at finite element boundaries based on a  traction-separation model. A major drawback of the CZM  is that crack patterns are highly sensitive to mesh orientation, unless extensive adaptive remeshing is used \cite{Tijssens2000,Zhou2004}.
The main reason why the CZM fails to produce mesh-independent results is that it is based on a traction-separation model that needs a priori knowledge of the potential crack surfaces. 
Even inserting a potential crack on each element interface is not enough 
to avoid mesh orientation dependencies (see for instance \cite{Rado2009}).
Another drawback  of the CZM is that the number of  degrees of freedom evolves in time because  nodes are doubled at separation (each element keeps a copy of the initially common node). This drawback disappears if all 
possible cohesive zone are pre-activated but this comes at the expense of a huge number of degrees of freedom.
Finally, note that the CZM approach is however very efficient when the crack path is known in advance
as shown in \cite{Lorentz2008b}.

A second major improvement of the Griffith model is the concept of diffuse or regularized damage modeling. The strength of diffuse damage models is to handle complex crack patterns with possible nucleation, branching and coalescence while providing results rather independent on mesh orientation. 
Diffuse damage models consist in a local stress-strain relation affected by a damage variable. The local model is then regularized with a length-scale to avoid spurious localization.
For the past thirty years, several types of 
regularization have been proposed in the 
literature  as the non-local integral damage model \cite{Pijau87,Lorentz2003a}; higher order, kinematically based gradient models \cite{aifantis_microstructural_1984,Schreyer86}
or  higher order, damage based, gradient models \cite{Fremond1996,Peerlings2001a,Nguyen2005b}.
Fracture was also recast in a regularized energy minimization problem  \cite{Francfort1998,Bourdin2000a} giving the so-called variational approach to fracture \cite{Bourdin2008}. 
At about the same time, the phase-field approach was
emanating from the physics community \cite{Karma2001} and then  developed for mechanics applications
\cite{Amor2009,Miehe2010,Kuhn2010,ambati_phase-field_2015}.
Finally, we can add the peri-dynamics approach \cite{Silling2000,PDReview2018} and the Thick Level Set approach \cite{Moes2011}.

This paper is about yet another diffuse approach to fracture based on a
Lipschitz regularization of the variable responsible for softening in the 
material model. It was introduced in \cite{Moes2021lipschitz}. 
For an elastic softening material, it requires
the damage variable field to be Lipschitz continuous.
It means roughly that the slope of the damage between any two points in the domain is bounded.
The Lip-field model is different from  gradient-damage or phase-field models because the expression of the free energy of the material  does not depend on the damage-gradient. It depends only on the strain and the damage (the classical local energy expression is basically kept).
The Lip-field model is however close in its conception to 
variational fracture or phase-field because it may be formulated as a minimization 
of an incremental potential to go from one time instant to the next.
The potential is identical to the one of the non-regularized (local) model. The idea of Lip-field is simply 
to enforce some specific regularity on the damage field.

Given the displacement and damage fields at some instant, 
a staggered scheme is used to find the fields at the next 
instant. The displacement update is a classical mechanical  problem with an imposed damage field. Under small strain and displacement assumptions, 
the problem can be linear or nonlinear depending on the stress-strain relation. 
We consider in this paper both symmetric and asymmetric tension-compression 
evolution for the damage. The former leads to a linear problem and the latter to a non-linear one.

As already described in \cite{Moes2021lipschitz}, 
the damage update in the Lip-field approach is rather different
than the one used in phase-field or damage gradient approaches. We still have to find the damage field as the minimizer of a convex function, but this time under Lipschitz constraints. Once discretized, this lead to a standard convex minimization problem under convex constraints that can be solved using standard packages. The computational efforts to solve this problem can be drastically reduced by taking advantage of some properties of the constraints. In particular, upper and lower bound to the damage field can be computed at very low cost, starting from a purely local (without the Lipschitz constraints) minimization. From these bounds, the size of the zone on which the non-local update needs to be computed (enforcing the Lipschitz constraints) is 
dramatically reduced.

An original aspect of the Lip-field implementation is that  damage irreversibly is automatically taken into account at no extra cost. Enforcing damage irreversibility is not straightforward for other type of diffuse approach as the phase-field for which it resorts to some approximation in the model. In \cite{Miehe2010b}, the variational inequality on damage growth is replaced by a variational equality in which the source term is replaced the maximum value of the source at previous times. The work \cite{Wheeler2014} is a rare work on phase-field in which the variational inequality is solved. 
Another original possibility offered by the Lip-field approach 
is to leave the damage variable at the integration points of the element  with the other internal variables. 
Thus, only displacement values are stored at the nodes.

The paper is organized as follows. 
The next section describes the classical damage mechanical formulation (non-regularized form) in the time-discrete setting. 
The Lipschitz regularization and its discretization is introduced in Section 3, followed by a discussion on an efficient strategy to construct the bounds mentioned above in Section 4.
Four examples of simulations are then detailed in Section 5, demonstrating the capability of the approach. A discussion of the results and  possible extensions to this work are discussed in the last section.

\section{The mechanical model: non-regularized formulation}
\label{sec:mecha}
We consider the deformation of a body, initially  occupying 
a domain $\Ome$, through a displacement field $\bfu$.
We assume small, quasi-static deformations. The Cauchy stress is denoted  $\bfsig$
and the strain  $\bfeps$ is given by the symmetric displacement gradient
\begin{equation}\label{key}
	\bfeps(\bfu) = \frac{1}{2} (\nabla \bfu + (\nabla \bfu)^\mathrm{T}) 
\end{equation}
where $\nabla$ indicates the gradient operator.
The displacement is imposed  on a part of the boundary denoted  $\Gam_u$ assumed fixed in time. 
On the rest of boundary,  zero traction forces are assumed (without loss of generality).
The set of kinematically admissible 
displacement fields at time $t$, 
is denoted $\cu(t)$:
\begin{equation}\label{eq:kin}
	\cu(t) = \{ \bfu \in H^1(\Ome): \bfu = \bfu_d(t) \texton \Gam_u \}
\end{equation}
where $\bfu_d(t)$ denotes the imposed displacement.
In the absence of body forces, the equilibrium condition reads
\begin{equation}\label{eq:stat}
	\int_{\Ome} \bfsig: \bfeps(\bfu^*) \dint \Ome  = 0, \quad  \forall \bfu^* \in \cu^*, \quad \cu^*= \{ \bfu \in H^1(\Ome): \bfu = 0 \texton \Gam_u \}
\end{equation}

Kinematics and equilibrium  equations (\ref{eq:kin}-\ref{eq:stat}) must be 
complemented with the constitutive model.
The formalism of generalized standard material introduced  in \cite{Halphen75,Germain83} is used.
The sole internal variable is the damage denoted  $d$. 
The model is characterized by a free energy potential $\varphi(\bfeps, d)$ and a 
dissipation potential $\psi(\dotd,d)$. 

We introduce an  implicit time-discretization and use the energetic variational approach. 
Given the displacement and internal variables $(\bfu_n, d_n)$ known at some instant $t_n$,  finding the pair 
$(\bfu_{n+1}, d_{n+1})$
at some later  instant $t_{n+1} = t_n + \Delta t$ amounts to a minimization problem  
\begin{equation}
	(\bfu_{n+1}, d_{n+1}) = \arg \min_{\substack{\bfu' \in \cu_n \\ d' \in \ca_n}} F(\bfu', d'; \bfu_n, d_n, \dt)
\end{equation}
where $\cu_n$ is a short-hand notation  for $\cu(t_{n+1})$ and 
$\ca_n$ enforces irreversibility and damage boundedness:
\begin{equation} 
	\ca_n = \{  d \in L^\infty(\Omega): d_n \leq  d \leq 1     \}  \label{eq:defD}
\end{equation}


For simplicity, we shall consider time-independent material models.
In this case, the $F$ expression does not depend explicitly on $\bfu_n, d_n$ and $\dt$. 
The extension to time-dependent models 
does not introduce difficulties. Also, to simplify the notations, 
we drop the $n+1$ indices. The non-regularized problem is then 
\begin{equation}
	(\bfu, d) = \arg  \min_{\substack{\bfu' \in \cu_n \\ d' \in \ca_n}} F(\bfu', d')  \label{eq:nonreg}
\end{equation}

The objective function is given as the integral over the domain of some local material objective function (and an extra term linear in $\bfu$ for non-zero body forces of surface tractions):
\begin{equation}
F(\mathbf u, d) = \int_{\Omega} f(\bfeps( \mathbf u), d) \dint \Ome
\end{equation}
The optimization problem it thus separable in $d$, meaning that knowing $\bfu$, finding $d$ is a local process at every point (this explains the qualification "internal" given to the $d$ variable).
The material
local objective function is composed of an strain energy term and a dissipation term:
\begin{equation}
\quad 
f(\bfeps( \mathbf u), d) = \varphi(\mathbf \bfeps( \mathbf u), d) + \Yc h(d)
\end{equation}
where $\Yc$ is the critical energy release rate and $h(d)$ is chosen as $h(d) = 2 d + 3 d^2$. The dissipation part is linked to the time integration of a dissipation potential given by 
$\psi(d, \dot{d}) = \Yc h'(d) \dot{d}$ where $h'$ is the derivative of $h$ with respect to $d$.
Regarding the free energy, we consider an assymetric tension-compression expression

\begin{equation}
	\varphi({\boldmath \epsilon},d)= \mu \sum_{i=1}^3 g(\alpha_i d) \epsilon_i^2 + \frac{\lambda}{2} g(\alpha d) \text{Tr}({\boldmath \epsilon})^2 
\end{equation}
where constants $\mu$ and $\lambda$ are the Lam\'{e} coefficients,  $\epsilon_i$, $i=1,2,3$, the eigenvalues of the strain tensor, $g(d)$ a convex function of $d$ describing the softening such as $g(0) = 1$ and $g(1) = 0$
and
\begin{equation}
\left\lbrace
\begin{array}{cccc}
\alpha_i = & \beta & \text{if} & \epsilon_i < 0       \\
           & 1     & \text{if} & \epsilon_i \geq 0   \\
\alpha   = & \beta & \text{if} & \text{Tr}({\boldmath \epsilon})   < 0       \\
           & 1     & \text{if} & \text{Tr}({\boldmath \epsilon})  \geq 0   
\end{array}\right.
\end{equation}
where $0 \leq \beta \leq 1$ is a user defined parameter. If $\beta = 1$, the behavior of the material is symmetrical in tension and compression. If $\beta = 0$, the material recovers its stiffness in compression and damage can only grow in tension. In the plane strain case ($\epsilon_3=0$) and for $\beta=1$, we can rewrite the energy as 
\begin{align}
\varphi(\mathbf \epsilon, d) = \phi_0(\mathbf \eps) + g(d)\phi_1(\mathbf \eps) + \Yc h(d)
\end{align}
with
\begin{align}
\phi_0 (\mathbf \epsilon) &= \begin{cases}
     0 &\text{if }\epsilon_1 \ge 0, \epsilon_2 \ge 0   \\
     \mu(\epsilon_1^2 + \epsilon_2^2) &\text{if } \epsilon_1 < 0, \epsilon_2 < 0  \\
     \mu\epsilon_1^2 & \text{if }    \epsilon_1 < 0, \epsilon_2 \ge 0  \\
         \end{cases} \\
\phi_1(\mathbf \epsilon) &= \frac{\lambda}{2} (\epsilon_1 + \epsilon_2)^2 + \begin{cases}
    \mu(\epsilon_1^2 + \epsilon_2^2) &\text{if }\epsilon_1 \ge 0, \epsilon_2 \ge 0   \\
    0 &\text{if }    \epsilon_1 < 0, \epsilon_2 < 0  \\
    \mu\epsilon_2^2 & \text{if }    \epsilon_1 < 0, \epsilon_2 \ge 0  \\
\end{cases} 
\end{align} 
where $\epsilon_1$ and $\epsilon_2$ are such that $\epsilon_1 < \epsilon_2$.
A common choice for the softening function $g(d)$ is to take $g(d) =(1-d)^2$. In this work, we used a generalized version:
\begin{align}
g(d) &= (1-d)^2 + \eta (1-d)d^3
\end{align}



\begin{figure}[ht]
	\begin{center}
		\input{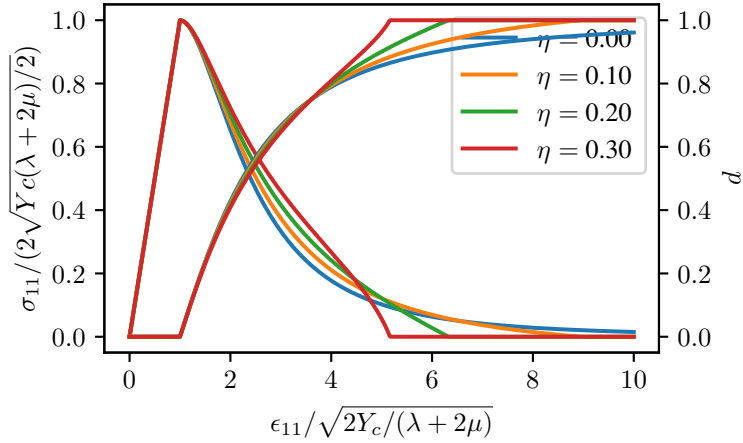}
	\end{center}
	\caption{Influence of the $\eta$ parameter on the stress/strain curve (left curves) 
	 and on the damage evolution with respect to strain (right curves).}
	\label{fig:g_eta}
\end{figure}

The parameter $\eta$ must be chosen in $[0, 1/3] $ to ensure convexity of $g$ for $d \in [0,1]$. The parameter $\eta$ allows to reach a damage of $1$ for a finite strain as indicated in figure \ref{fig:g_eta}.
Damage starts to grow for a strain  $\sqrt{\frac{2 Yc}{\lambda + 2\mu}}$ independent of $\eta$ whereas
damage reaches $1$ for a strain of $4\sqrt{\frac {\Yc}{  (\lambda + 2\mu)/\eta}}$.
One can easily check that $f$ (and thus $F$) is convex separately in $\bfu$ and $d$, but not in both, giving the well known damage softening effect.


 
\section{Lipschitz regularization and discrete setting}

The main idea of the Lip-field approach is to impose a Lipschitz  
regularity condition on the damage field. The regularity set is defined by 
\begin{equation}
	\Lip = \{ d \in L^{\infty}(\Ome): 
	\abs{ d(\bfx) - d(\bfy)}  \leq \frac{1}{l} \, \dist(\bfx, \bfy), \quad \forall \bfx, \bfy \in \Ome \} \label{eq:lip}
\end{equation}
where $l$ is the regularizing length and $\dist(\bfx,\bfy)$ is the minimal length of the path inside $\Ome$ joining $\bfx$ and $\bfy$ (the distance is considered infinite if the two points cannot be connected inside $\Ome$). 
The Lipschitz regularized problem  is
obtained by adding the Lipschitz constraint to the 
non-regularized problem \eqref{eq:nonreg}
\begin{equation}
	(\bfu, d) = \arg  \min_{\substack{\bfu' \in \cu_n \\ d' \in \ca_n \cap \Lip}} F(\bfu', d')  \label{eq:reg}
\end{equation}
On the contrary to \eqref{eq:nonreg},  the  above optimization  is no longer separable in $d$ because the Lip-constraint ties spatially damage. 
It is however still convex $\bfu$ and $d$ separately (since the set $\Lip$
is convex).

\subsection{Spatial discretization}
\begin{figure}[ht]
	\begin{center}
		\includegraphics[width=0.7\textwidth]{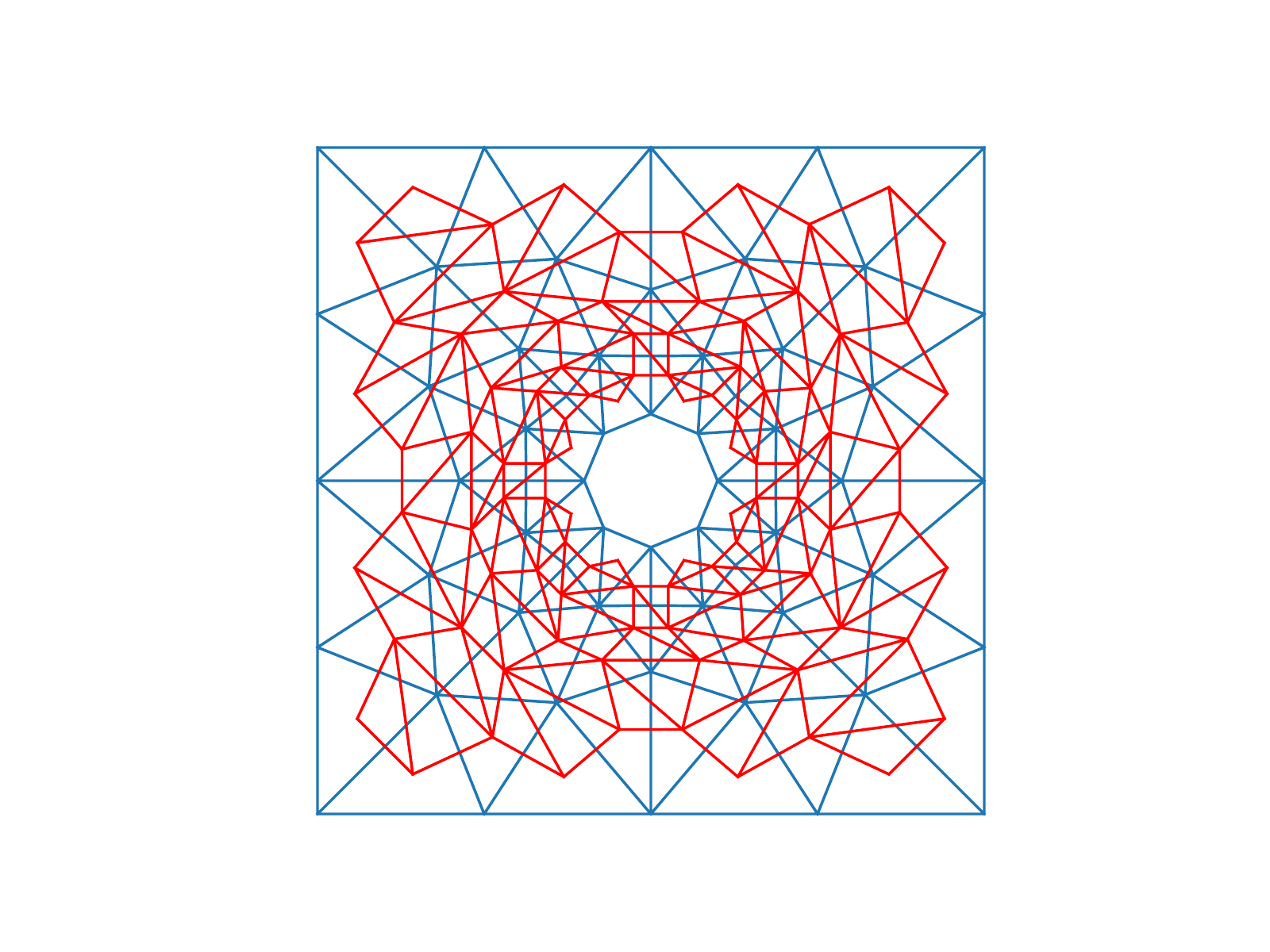}
	\end{center}
	\caption{The finite element mesh (blue) and  the Lip-mesh (red) built from the centroids of the blue mesh elements.}
	\label{fig:lipmesh}
\end{figure}

The domain $\Omega$ is discretized into a geometrical mesh denoted
$\Omeh$. An example is depicted in blue for a plate with a hole in Figure \ref{fig:lipmesh}. 
We then consider a classical finite element discrete space on $\Omeh$.
The displacement is linear over each element and continuous over the mesh. The strain is thus piecewise constant.
The damage is stored at the centroid of each element (classical finite element approach for internal variables).
To express the Lip constraint, we build an additional triangular mesh, called Lip-mesh and denoted $\Delta^h$, linking the centroids of the elements (red mesh in Figure \ref{fig:lipmesh}).  This mesh is built once and for all.
The set of vertices, edges and elements of $\Delta^h$ are denoted $V$, $E$ and $T$, respectively. The  Lip-mesh is embedded inside the displacement mesh 
($\Delta^h \subset \Omeh$). The domains covered by 
$\Delta^h$ and  $\Omeh$ have the same topology. 
The Lip-mesh does not add new holes and complies with the hole of the displacement mesh. 
The damage field is discretized in a piecewise linear 
continuous fashion over the Lip-mesh. The discrete damage space is denoted $D^h(\Delta^h)$.  
The damage gradient is thus piecewise constant on the Lip-mesh. 

The continuum Lipschitz set \eqref{eq:lip} involves an infinite number of constraints since all pairs of points must be considered. We need to find a way to discretize this set. 
A first option is to bound the damage gradient on every element of the 
Lip-mesh \eqref{eq:liph}. A second option is to enforce the Lipschitz 
constraint in between vertices (\ref{eq:liphu}-\ref{eq:liphd}) .
The metric $\disth(\bfx, \bfy)$  is the shortest path between $\bfx$ and $\bfy$   lying inside $\Delta^h$, whereas the discrete metric $\disthp(\bfx, \bfy)$ forces the path to go along edges 
(and thus $\disth(\bfx, \bfy) \leq  \disthp(\bfx, \bfy)$).
Finally, another option is to  enforce the Lipschitz constraint on each edges
\eqref{eq:liphp}.
\begin{align}
	\Liph & = \{ d \in D^h(\Delta^h):  \normgrad{d} \mid_t  \leq \frac{1}{l}, \quad  \forall t \in T \}\label{eq:liph}  \\
	\Liphu & = \{ d \in D^h(\Delta^h):  \abs{ d(\bfx) - d(\bfy)}  \leq \frac{1}{l} \, \disth(\bfx, \bfy), \quad \forall \bfx, \bfy \in V \} \label{eq:liphu}\\
	\Liphd & = \{ d \in D^h(\Delta^h):  \abs{ d(\bfx) - d(\bfy)}  \leq \frac{1}{l} \disthp(\bfx, \bfy), \quad \forall \bfx, \bfy \in V \} \label{eq:liphd}\\
    \Liphp & = \{ d \in D^h(\Delta^h):  \abs{ d(\bfx) - d(\bfy)}  \leq \frac{1}{l} \norm{\bfx-\bfy}, \quad \forall (\bfx, \bfy) \in E \} \label{eq:liphp}
\end{align}
The four options satisfy the following inclusions proven in  appendix \ref{appendix:incl}:
\begin{equation}
	\Liph  \subset \Liphu \subset \Liphd = \Liphp
	\label{eq:inclusions}
\end{equation}

We choose the first option because it involves the least number of discrete constraints since the number of element in a mesh is much smaller than the number of edges (or pairs of vertices). Also, compared to $\Liphp$ it is less prone to mesh orientation effect because we are checking the Lipschitz constraint on all orientations 
and not only along along edges (see appendix \ref{appendix:exampleproj}).
The space-time discrete problem is thus to find 
at time $t_{n+1}$, the pair $(\bfu, d)$ satisfying 
\begin{equation}
	(\bfu, d) = \arg  \min_{\substack{\bfu' \in \cu^h_n \\ d' \in \ca^h_n \cap \Liph}} F(\bfu', d')  \label{eq:optdis}
\end{equation}
where $\cu^h$ indicates the displacement finite element 
space and $\ca^h_n$ is given by 
\begin{equation} 
	\ca^h_n = \{  d \in D^h(\Delta^h): d_n(\bfx) \leq  d(\bfx)  \leq  1, \forall \bfx \in  V    \}  
\end{equation}

\subsection{Staggered scheme}
The optimization problem 
\eqref{eq:optdis} is not convex 
with respect to the pair 
$(u,d)$ but is convex with respect to each variable taken separately.
As for the phase-field approach \cite{Miehe2010b}, 
a staggered scheme is used: solve for the displacement  with given  damage, then solve for  damage with given displacement  and iterate until convergence. 
The first minimization is rather standard. 
The second one is not. It reads
\begin{equation}
	d = \arg  \min_{\substack{d' \in \ca^h_n \cap \Liph}} F(\bfu, d')  \label{eq:dsought}
\end{equation}
where $\bfu$ is  known  (current iterate).
It is a convex minimization problem with cone and second order cone inequality. To solve it we use the \textit{cvxopt} \cite{cvxopt} package, in particular the \textit{cp} function that can find the minimizer of a general convex function with first and second order cone constraints. An example of Lipschitz projection is given in  appendix \ref{appendix:exampleproj} for a constructed damage field.

\section{The use of bounds for the damage update}
The previous section did detail the space discretization 
as well as the staggered scheme. The damage iterate consists in a convex optimization. It has been observed in the first Lip-field paper \cite{Moes2021lipschitz} that this 
optimization could be greatly simplified using so-called bounds on the  solution. Below, we recall the bound concept and detail how it is implemented  in the discrete setting.

\subsection{Bounds in the continuum setting}
Bounds on the damage solution $d$ have been proposed in the paper \cite{Moes2021lipschitz}. 
Consider the damage optimization problem 
\begin{equation}
	d = \arg  \min_{\substack{d' \in \ca_n \cap \Lip}} F(\bfu, d')  
	\label{eq:optid}
\end{equation}
The idea is to first disregard the Lipschitz constraint and compute a local damage update denoted  $\dloc$:
\begin{equation}
	\dloc = \arg  \min_{\substack{d' \in \ca_n}} F(\bfu, d')  \Rightarrow
	\dloc(\bfx) = \arg  \min_{\substack{d_n(\bfx) \leq d' \leq 1}} f(\bfeps(\bfu)\mid_{\bfx}, d'), \quad \forall \bfx \in \Ome
\end{equation}
In the above, we have use the separability property of the 
non-regularized optimization with respect to the $d$ variable. The local damage at point $\bfx$ only depends on the strain at that point.
Upper and lower bounds are defined as 
\begin{align}
	\dl(\bfx) & = \min_{\bfy \in \Ome} \; (\dloc(\bfy) + \frac{1}{l} \dist(\bfx, \bfy)) \label{eq:lowerbound} \\
	\du(\bfx) & = \max_{\bfy \in \Ome} \;  (\dloc(\bfy) - \frac{1}{l} \dist(\bfx, \bfy)) \label{eq:upperbound}
\end{align}
They satisfy
\begin{equation}
	d_n \leq \dl \leq \dloc \leq \du \leq 1, \quad 
	\dl \leq d \leq \du  \label{eq:ineq}
\end{equation}
The proof may be found in  \cite{Moes2021lipschitz}. 
The optimization for $d$ in \eqref{eq:optid} may thus be replaced by
\begin{equation}
	d = \arg \min_{\substack{d' \in \oca_n \cap \Lip }} F(\bfu, d')
\end{equation}
where 
\begin{equation}
    \oca_n = \{  d \in L^\infty(\Omega): \dl  \leq  d \leq \du    \}
    \subset \ca_n
\end{equation}
It is clear  that at any point for which the bounds are equal, the local damage update is optimal:
\begin{equation}
    \dl(\bfx) = \du(\bfx) \Rightarrow d(\bfx) = \dloc(\bfx)
\end{equation}

The bounds computation may thus potentially drastically reduce the effort for 
the damage optimization by locating the subdomain over which the local update needs to be further corrected.
The subdomain is an upper-bound for the zone over which the Lipschitz constraint  will be active.

\subsection{Bounds in the discrete setting}
First, a local update may be computed at each vertex 
\begin{equation}
	\dloc(\bfx) = \arg  \min_{\substack{d_n(\bfx) \leq d' \leq 1}} f(\bfeps(\bfu)\mid_{\bfx}, d'), \quad \forall \bfx \in V  
	\label{eq:dloc}
\end{equation}
Then, we propose to compute the following bounds
\begin{align}
	\dl(\bfx) & = \min_{\bfy \in V} \; (\dloc(\bfy) + \frac{1}{l} \disthp(\bfx, \bfy)), \quad  \forall \bfx \in V \label{eq:bl} \\
	\du(\bfx) & = \max_{\bfy \in V} \;  (\dloc(\bfy) - \frac{1}{l} \disthp(\bfx, \bfy)), \quad \forall \bfx \in V   \label{eq:bu}
\end{align}
The definition above seems to indicate that $O(n^2)$ operations are required, where $n$
is the number of vertices in $V$. 
Fortunately, an algorithm inspired from Dijkstra's algorithm \cite{Dijkstra59} reduces the effort to 
$O(n \log n)$ operations. It is detailed in the appendix \ref{appendix:dijkstra}. At any vertex, the bounds satisfy similar inequalities than the bounds in the continuum setting (see proof in the appendix \ref{appendix:proof_bounds}):
\begin{equation}
	d_n(\bfx) \leq \dl(\bfx) \leq \dloc(\bfx) \leq \du(\bfx) \leq 1, \quad 
	\dl(\bfx) \leq d^+(\bfx) \leq \du(\bfx), \quad \forall \bfx \in V
	\label{eq:disbound}
\end{equation}
where 
\begin{equation}
	d^+ = \arg \min_{\substack{d' \in \ca^h_n \cap \Liphp }} F(\bfu, d') \label{eq:dplus}
\end{equation}

Unfortunately, the bounds do not bracket the solution $d$ from we are after \eqref{eq:dsought}  but another solution defined by \eqref{eq:dplus}.
The reason for this discrepancy is that we have chosen the metric $\disth$ instead of $\dist$ to 
compute the bounds because the former gives an extremely simple algorithm.
To take into account the discrepancy, we use the following scheme.
\begin{itemize}
	\item Step 1: Compute the bounds \eqref{eq:bl}-\eqref{eq:bu} using the algorithm given in the appendix \ref{appendix:dijkstra}.
	\item Step 2:  Initialize $V^l$ to the set of vertices for which  $\dl(\bfx) = \du(\bfx)$ and define $T^l$ as the set of elements connected to nodes only in $V^l$.
	\item Step 3: Solve 
	 \begin{equation}
	 	d = \arg \min_{\substack{d' \in \oca^h_n \cap \Liphb }} F(\bfu, d') 
	 \end{equation}
	 where 
	 \begin{align}
	 	\oca^h_n & = \{  d \in D^h(\Delta^h):  d(\bfx) = \dloc(\bfx), 
	 	\forall \bfx \in V^l \text{\ and\ }    d_n(\bfx) \leq d(\bfx) \leq 1, \forall \bfx \in V \setminus V^l    \} \\
	 	\Liphb  & = \{ d \in D^h(\Delta^h):  \normgrad{d} \mid_t  \leq \frac{1}{l}, \quad  \forall t \in T \setminus T^l \} 
	 \end{align}
    \item Step 4: If $d \in \Liph$ end, else remove the element for which $\normgrad{d} > 1/l$ from $T^l$ and remove its node from $V^l$, go to Step 3.
\end{itemize}

If the bounds were associated to the true solution, step 4 would end directly. In the numerical 
experiments  we have noticed that Step 3 was repeating itself only on rare occasions, showing 
that the discrepancy is not too detrimental.

\section{Simulation results}
All examples are computed using plane strain assumption. The meshes and the python scripts used for are all the examples are open-source and may be downloaded from \href{https://gitlab.com/c4506/lipfield}{https://gitlab.com/c4506/lipfield}.
All examples are treated with the symmetrical 
model ($\beta =0$) except for the shear test 
which is treated with both the symmetrical 
($\beta =0$) and asymmetrical ($\beta = 1$) models.

For all examples, 
the mesh is built using the open-source gmsh software \cite{geuzaine2009gmsh}. Regarding the Lip-mesh, it is constructed using the open-source Triangle software \cite{Shewchuk1996} available 
at \url{https://www.cs.cmu.edu/~quake/triangle.research.html} along 
with a python interface available at \url{https://github.com/pletzer/pytriangle}.

\subsection{Plate with a hole}
\begin{figure}[ht]
	\begin{center}
		\includegraphics[width=1.\textwidth]{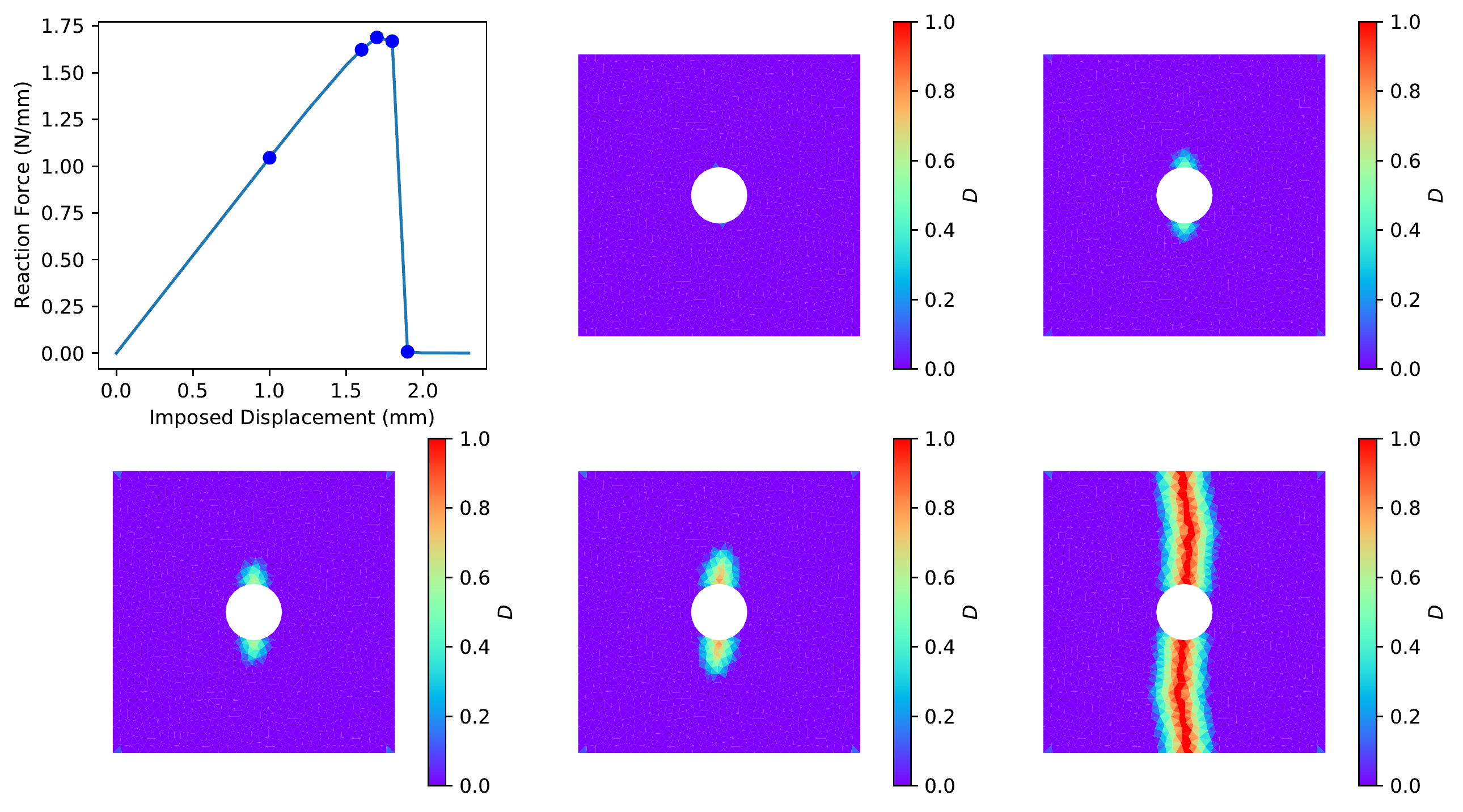}
	\end{center}
	\caption{Load displacement curve and damage field for mesh 2.}
	\label{fig:holeplate}
\end{figure}

As a first example, we pull the sides of a square domain with imposed displacement. The square is pierced at its center by with a circular hole. The geometric and material parameters used are reported on table \ref{tab:holeinplategeompar}. The material parameters are chosen generically for this first example since  we are just trying here to assess the method qualitatively.
\begin{table}[ht]
	\begin{center}
		\begin{tabular}{cc}
		\begin{tabular}{|r|r|r|}
			\hline
			units  &$L$ & $R$ \\
			\hline
			mm &2 &0.2 \\
			\hline  
		\end{tabular} &
	
	\begin{tabular}{|l|l|l|l|l|}
		\hline
		$E$ & $\nu $ & $l$ & $\Yc$ & $\eta$ \\
		\hline
		1 MPa &	0.2 & 0.2 mm & 1 MPa & 0.1\\
		\hline  
	\end{tabular}
    \end{tabular}
	\end{center}
	\caption{Hole in a plate:  geometrical parameters (left), material parameters (right).}
	\label{tab:holeinplategeompar}
\end{table}
The mesh is unstructured and parametrized by the size of $h$ the edge of the triangles on the boundary of the square and the circular hole.
Results are reported on figure \ref{fig:holeplate}. On the top left, we have the load/displacement curve. We have first an elastic loading, until the stored energy is sufficient to trigger damage. We then have a short softening phase, where the damage start to grow gently at two symmetric locations at the top and bottom of the hole, followed by an abrupt release of strain energy during which the damage zones develop in two cracks propagating up to the boundary of the plate. This setting clearly shows a strong instability, and the equilibrium path is discontinuous under displacement loading. To improve on that, we could consider another control mechanism in order to follow the snap-back behavior.





\begin{table}[ht]
	\begin{center}
		\begin{tabular}{|l|r|r|r|}
			\hline
			mesh  & $L/h$ & vertices & faces \\
			\hline
			\hline
			mesh 1      & 16 & 1524& 2896\\
			lip-mesh 1  &  & 2896& 5476\\
			\hline
			\hline 
			mesh 2      & 32 & 5640  &10980\\
			lip-mesh 2  &  & 10980 &21308\\
			\hline
			\hline 
			mesh  3     & 64 & 19333  &38070\\
			lip-mesh 3  &  & 38070 &74947\\
			\hline
			
		\end{tabular}
	\end{center}
	\caption{Mesh  parameters for the convergence analysis.}
	\label{tab:holeinplatemeshpar}
\end{table}

\begin{figure}[ht]
	\begin{center}
		\includegraphics[width=1.\textwidth]{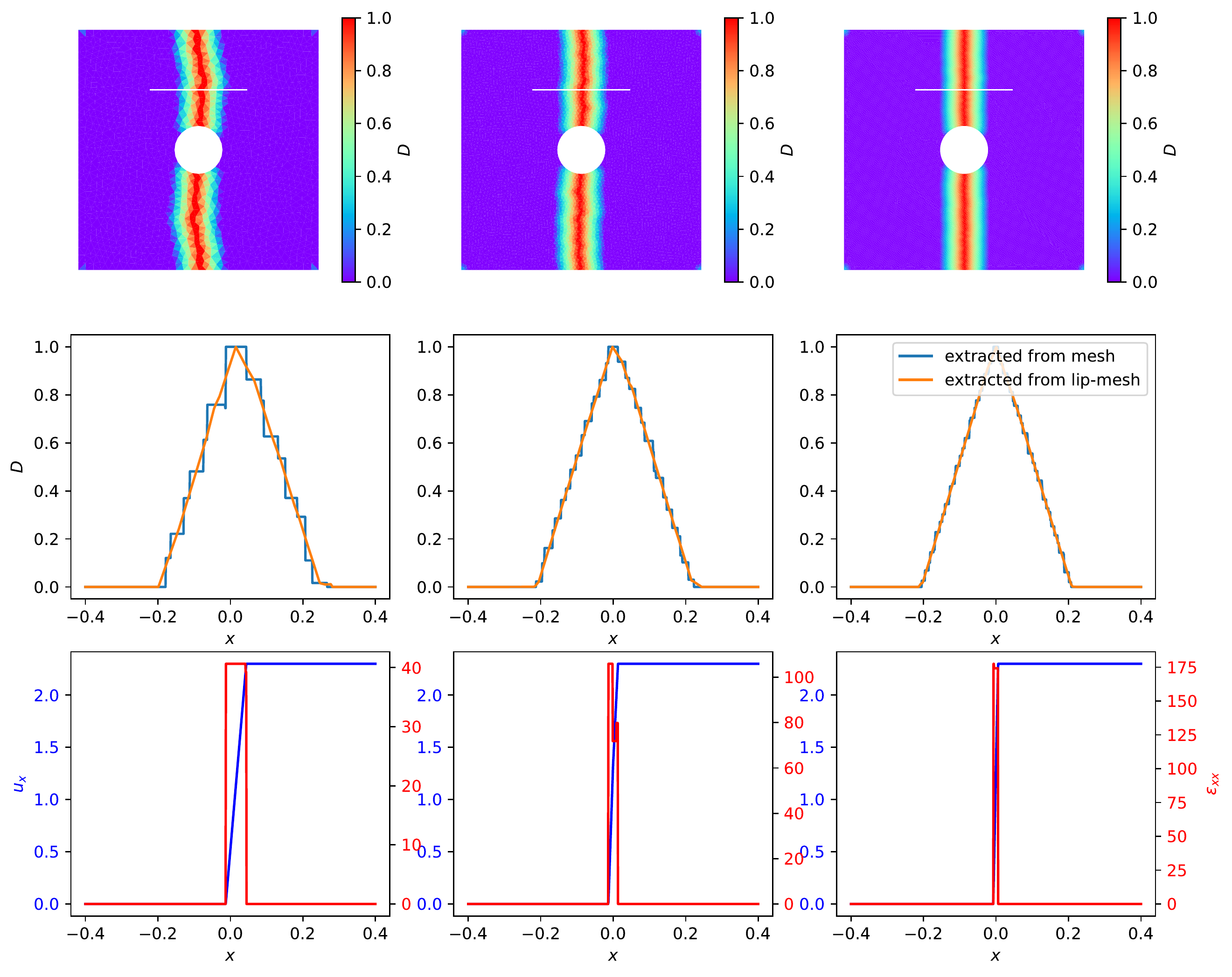}
	\end{center}
	\caption{Convergence analysis: top damage field, below $u_x$, $\epsilon_{xx}$ extracted from the white line for different meshes, at the end of the simulation.}
	\label{fig:holeplateconv}
\end{figure}

Figure \ref{fig:holeplateconv} reports on a convergence analysis. All the parameters are kept identical except for the mesh density which is increased as show on table \ref{tab:holeinplatemeshpar}. The damage field is plotted on the top row, with the finer mesh on the right.  The two damage zones take the shape of a vertical band, of thickness close to $2l$ where the damage reach $1$ at the center of the band. The elements for which $d$ reaches $1$ have of course no stiffness and could be identified as cracks. On the coarsest mesh, the band is not perfectly straight due to the unstructured nature of the mesh. As the mesh is refined, while keeping $l$ constant, more and more elements fit into the thickness of the band which appears straighter and straighter while the damage field is described with more and more precision. On the middle row of the figure, we report the damage field, for each mesh, along the white line represented on the top row. The white line is perpendicular to the crack, and the damage field clearly reaches $1$ on the crack, and reduces at a slope fixed by the Lipschitz constraint away from the crack. The thickness of the damaged zone clearly converges quickly toward the value of $2l$. We plot for each mesh the damage field as seen by the mechanical problem (constant per mesh element) and the damage field as seen by the Lip-field problem (linear on each element of the Lip-Mesh). Note that both fit very well to the expected wedge shape imposed by the Lip-field constraint. This means that in absence of the constraint, the damage would be zero everywhere away from a strip of 1 element thickness, reproducing the classical mesh dependency of the result for a local damage model. The last row of the figure reports, along the same line, the displacement and the strain field. Notice how the displacement clearly displays a sharp jump over the thickness of a single element, reproducing the result that would have been obtained if we would have inserted a sharp crack in the mesh. The deformation component $\epsilon_{xx}$ is also plotted and the expected Dirac-like function shape is captured.

\subsection{Comparison with a Griffith analysis}
 
\begin{figure}[ht]
	\begin{center}
		\def\svgwidth{0.6\textwidth}
		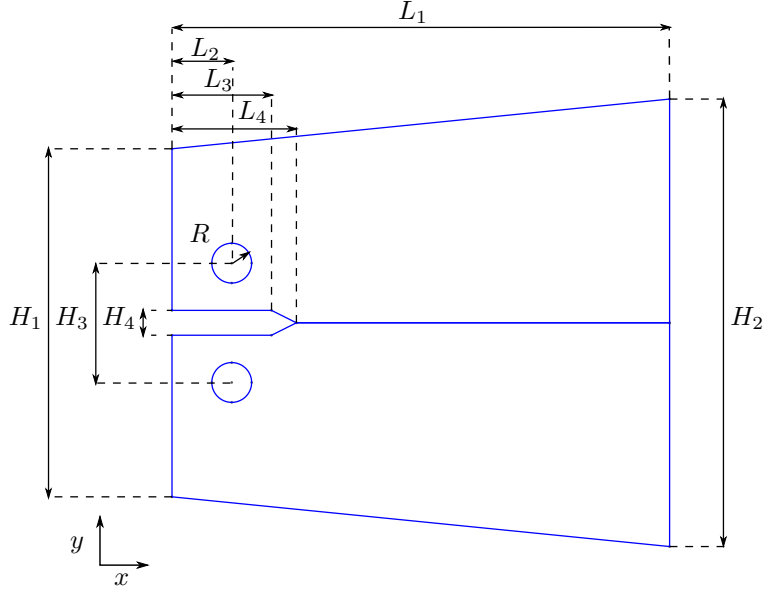
	\end{center}
	\caption{TDCB geometry definition.}
	\label{fig:tdcbgeom}
\end{figure}

The geometry described in figure \ref{fig:tdcbgeom} is used to study crack propagation.
This geometry is inspired from the classic TDCB  (tapered double cantilever beam) shape that insures a stable crack growth relative to an increasing displacement loading.

\begin{table}[ht]
	\begin{center}
		\begin{tabular}{|l|c|c|c|c|c|c|c|c|c|}
			\hline
			units  &$L_1$ & $L_2$ & $L_3$ & $L_4$ & $H_1$ & $H_2$ & $H_3$ & $H_4$ & $R$ \\
			\hline
			mm &100 &12 &  20 & 24 & 70 &  90 & 24 & 5 & 4 \\
			\hline  
		\end{tabular}
	\end{center}
	\caption{TDCB geometrical parameters.}
	\label{tab:tdcbgeompar}
\end{table}

\begin{table}[ht]
	\begin{center}
		\begin{tabular}{|c|c|c|c|}
			\hline
			$E$ & $\nu$ & $K_{I_c}$  & $\eta$ \\
			\hline
			3500 $\mathrm{MPa}$& 0.32 &  1.4 $ \mathrm{MPa\sqrt m}$  & 0.1 \\
			\hline  
		\end{tabular}
	\end{center}
	\caption{TDCB material parameters.}
	\label{tab:tdcbmatpar}
\end{table}

The dimensions defining the geometry are given in table \ref{tab:tdcbgeompar}.
The TDCB specimen is loaded by a linearized rigid body motion on the boundary of the two holes. The bottom hole has its center fixed, while it is free to rotate around the $z$ axis. The top hole is free to rotate around the $z$ axis, its center is fixed on the $x$ axis, while its motion on the $y$ axis is imposed and is the loading parameter $u$. We make the classical plane-strain and small-strain assumption. The material is described with the coefficients given in table \ref{tab:tdcbmatpar}, where $E$, $\nu$ and $K_{I_c}$ denote respectively, the Young Modulus, Poisson ratio and the critical mode I stress intensity factor.
Upon loading, a crack is expected to develop at the notch tip, and propagate along the middle axis of the TDCB specimen, which will be called the crack path. For a first validation of the Lip-field approach for brittle damage we will compare our results with a linear elastic fracture mechanics analysis, using a Griffith criteria.
\paragraph{Griffith Analysis}
According to the Griffith criterion, the crack should propagate if the elastic energy release rate per unit of crack length $G$ is equal to $\Gc$, where $\Gc$ is a material parameter: the critical energy release rate. Under plane strain assumption, for isotropic-elastic material, we have
\begin{equation}
\Gc = \frac{1-\nu^2}{E}{K_{I_c}}^2
\end{equation}
In order to compute the load-displacement ($F(u)$) and the crack-length ($a(u)$) under Griffith assumption, we applied the following approach. The mesh is constructed in two symmetrical-part along the crack-path, so that the nodes on the crack-line are regularly spaced and we denote by $h$ the spacing between two nodes. The mesh of the upper part and the mesh of the lower part are disconnected, but the nodes of each part are at the same geometrical positions on the crack-path. The nodes are connected using a Lagrange multiplier to enforce the same displacement on both part at each of these node, on the section of the crack-path  to the right of the actual crack. With this setting, we can easily compute the equilibrium displacement field $\mathbf u$ and the strain energy $e$ in the TDCB specimen using the standard finite element method, for different crack-length and a unit displacement $u$. For each crack length $a$, corresponding to a number of unconnected nodes on the crack path, we can compute the energy release-rate $G_1(u)$ for the unit displacement, by using a first order Taylor expansion:
\begin{equation}
	G_1(a) = \frac{e_1(a+h) - e_1(a-h)}{2h}
\end{equation}
where $e_1(a)$ is the strain energy at equilibrium for the crack of length $a$ and a imposed displacement $u=1$. Since $G$ is clearly a quadratic function of $u$, we can compute for each value of $a$ the value of $u_c$ and then $f_c$ for which $G=G_c$, and plot the $F(u)$ and $a(u)$ curves for the case of Griffith analysis. 

\paragraph{Lip-field Analysis}
In order to compare Griffith analysis and Lip-field, we need to set up the parameter of the damage model in order to be energetically equivalent to Griffith. Assuming that when the damage localize into a shape of a crack, and that $d$ saturates the Lipschitz constraint and reaches $1$ on the crack path. We can approximate the $d(y)$ function  on a line perpendicular to the crack path with a triangular profile of slope $1/l$: 
$d(y) = \frac {\max(l-\abs{y},0)}{l}$, where $y$ is  the distance to the crack path.  We can then evaluate the energy dissipated per unit of crack advance by $\Gc = \Yc \int_{-l}^{l} h(d(y)) dy = 4\Yc l$.
In the simulations, we set $l$ so that we have enough elements in the band to represent the crack (typically $l = 5h$), and then set $\Yc = \Gc/(4l)$.

\begin{figure}[ht]
	\begin{center}
		\includegraphics[width=1.\textwidth]{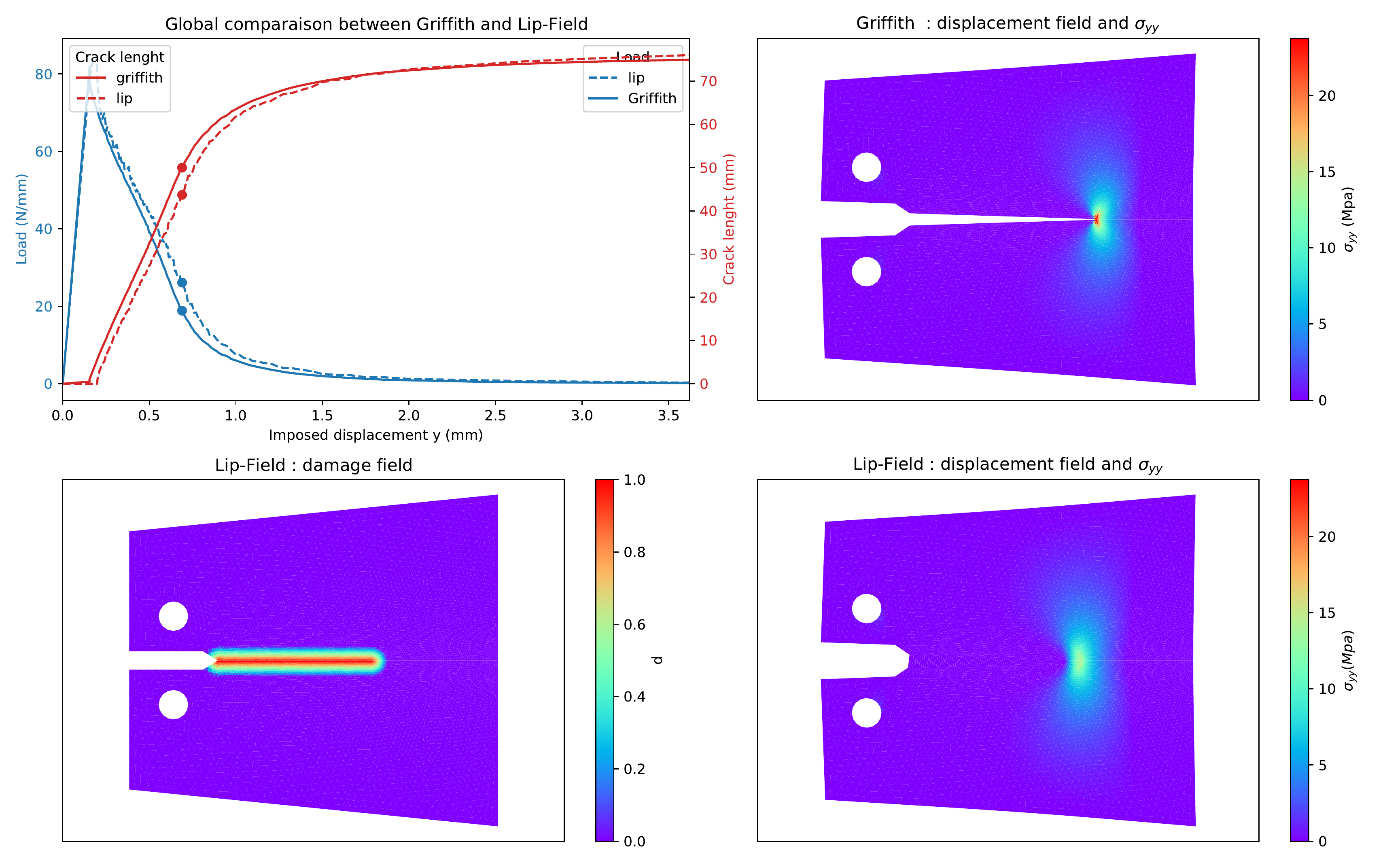}
	\end{center}
	\caption{Griffith/Lip-field comparison.}
	\label{fig:griflip}
\end{figure}

The comparison between Griffith and Lip-field is given on figure \ref{fig:griflip} where we report the reaction on the anchoring circle and the crack length as a function of the imposed displacement, for both Griffith and Lip-field analysis. 

In both cases, after reaching a critical displacement, the reaction drops in a controlled way, while the crack advance at a pace proportional to the imposed displacement pace, until the crack length cover roughly three quarters of its maximum length. After that, the pace of crack progressively slows down, until it reaches the right boundary of the sample. The results compare well:  we observe the same critical displacement, and the same speed between Griffith and Lip-field, even if we have a slight difference in dissipated energy and crack tip position. A more extensive study is beyond the scope of this paper, but we have already a good confirmation of the ability of Lip-field to reproduce Griffith physics.


\subsection{Shear test}
\begin{figure}[ht]
	\begin{center}
		\def\svgwidth{\textwidth}
		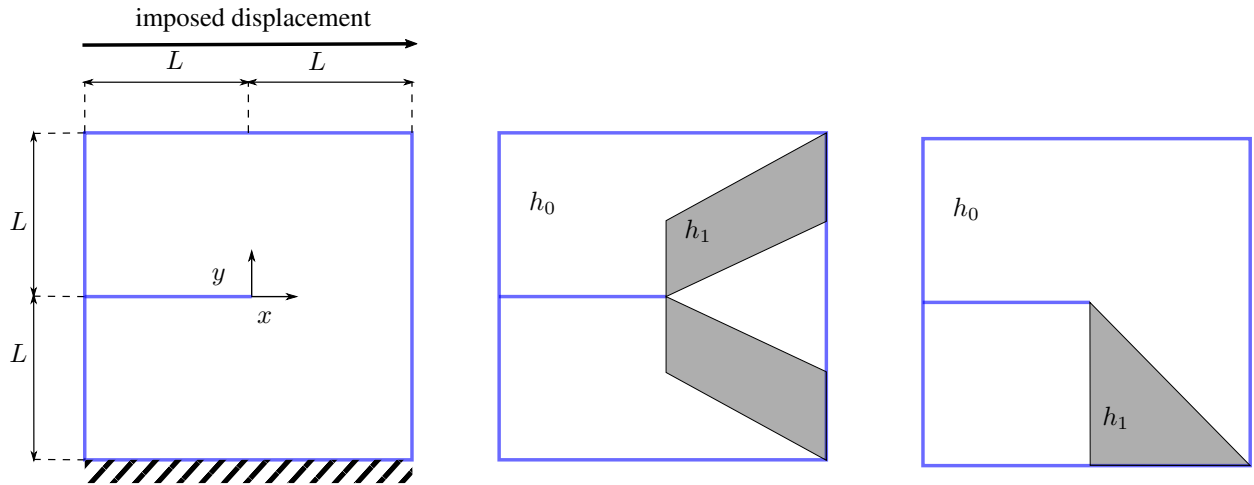
	\end{center}
	\caption{From left to right : shear test geometry and boundary conditions definition, mesh parameter for the symmetric case and the asymmetric case}
	\label{fig:sheartestgeom}
\end{figure}

In the next example, we perform a shear test on the geometry given in figure  \ref{fig:sheartestgeom}.  A square sample of length $ 2L = 1 \mathrm{mm}$ is initially cut by a line, from the left boundary to the center. This is meant to represent an initial crack. The bottom of the square is fixed, while the top has its displacement fixed to 0 in the $y$ direction and imposed in the $x$ direction.  This benchmark  is popular in the phase-field literature \cite{wu2020phase}.
 The material properties are given in table \ref{tab:sheartestmatpar}, and $\Yc$ is computed to obtain energetic equivalence as discussed in previous subsection ($\Yc = G_c/(4l)$). The parameter $\eta$ is set to $0.1$. The two parameters  $h_0$ and $h_1$ are  the target edge sizes for the 
 meshing tool  in the coarse and fine zone, respectively, as indicated in figure \ref{fig:sheartestgeom}. The extent of the fine zone has been determined thanks to an initial run on a coarse mesh in order to identify the zone in which the damage localizes into a crack. The parameters for the two meshes used for this example are given in table 
\ref{tab:meshparshear}.

\begin{table}[ht]
	\begin{center}
		\begin{tabular}{|c|c|c|c|c|}
			\hline
			$E$ & $\nu$ & $l$ & $G_c$ & $\eta$ \\
			\hline
			210000 $\mathrm{MPa}$& 0.2 &  0.015$\mathrm{mm}$ & 2.7 $\mathrm{MPa/mm}$ & 0.1 \\
			\hline  
		\end{tabular}
	\end{center}
	\caption{Shear test material parameters.}
	\label{tab:sheartestmatpar}
\end{table}

\begin{table}[ht]
	\begin{center}
	\begin{tabular}{|l|r|r|r|r|}
		\hline
		mesh  & $h_0$ & $h_1$ & vertices & faces \\
		\hline
		\hline
		symmetric & 1/8 mm & 1/128 mm  & 26656& 53311\\
		\hline 
		asymmetric   & 1/8 mm  & 1/128 mm & 11343  &22685\\
		\hline
		\hline 
	\end{tabular}
\end{center}
\caption{Mesh parameters for the shear test.}
\label{tab:meshparshear}
\end{table}

\begin{figure}[ht]
	\begin{center}
		\includegraphics[width=1.\textwidth]{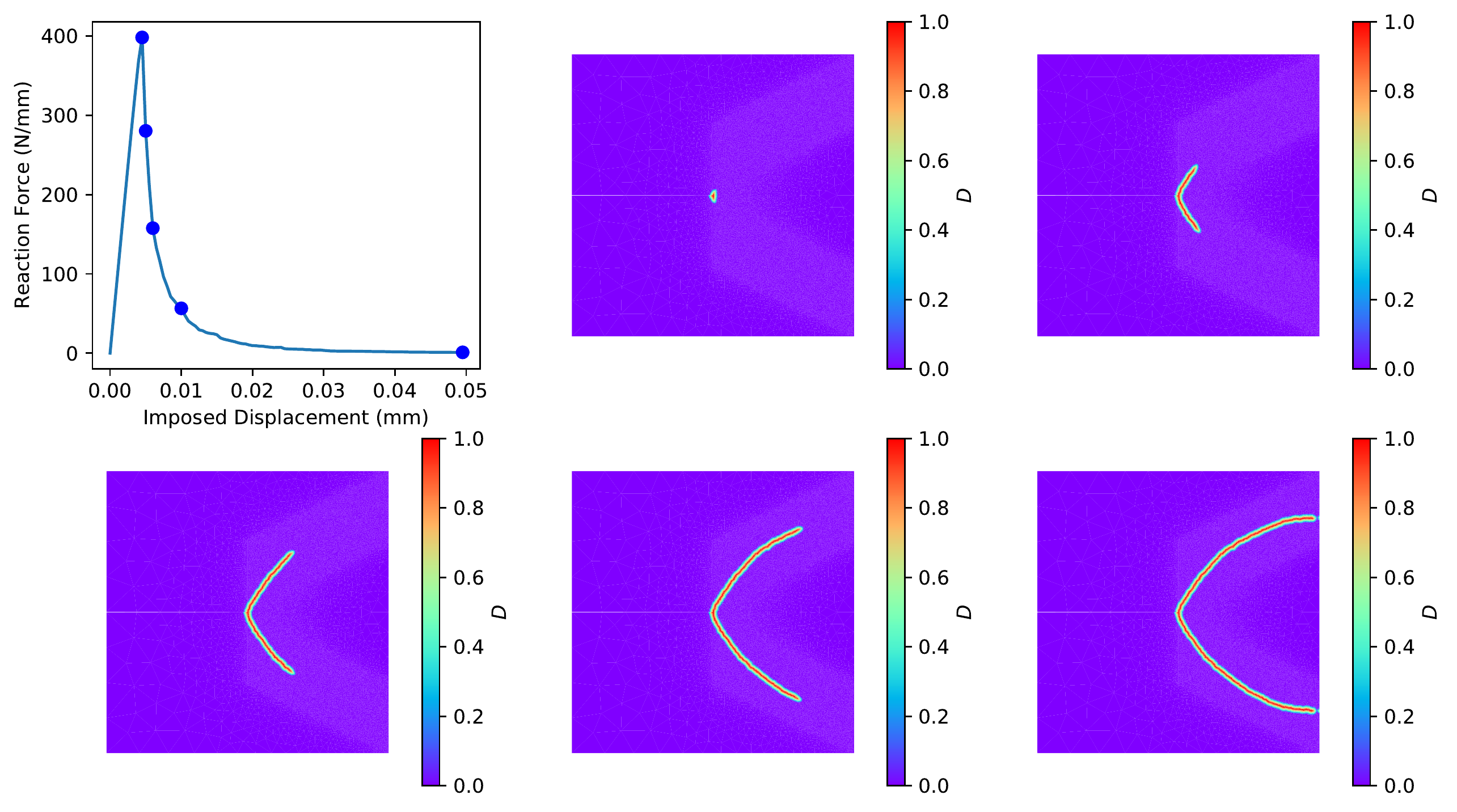}
	\end{center}
	\caption{Shear test: symmetric traction/compression. Top left: load displacement curve. From left to right and top to bottom, damage field at the corresponding dot on the curve.}
	\label{fig:sheartestld}
\end{figure}

\begin{figure}[ht]
	\begin{center}
		\includegraphics[width=1.\textwidth]{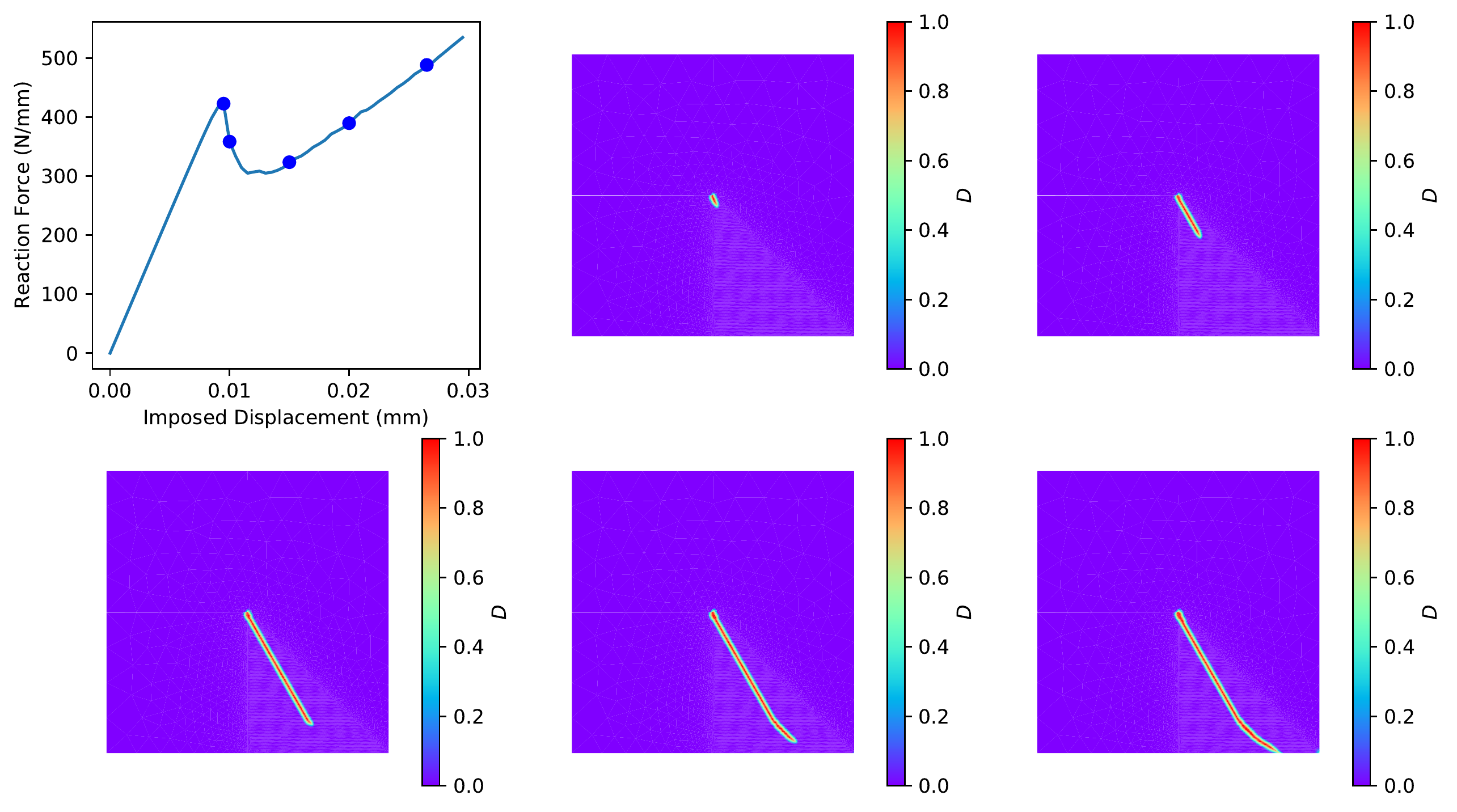}
	\end{center}
	\caption{Shear test: Asymmetric traction/compression. Top left: load displacement curve. From left to right, top to bottom damage field at the corresponding dot on the curve.}
	\label{fig:sheartestld_assym}
\end{figure}

We report on figure \ref{fig:sheartestld}, the load/displacement curve, as well as the damage field, at different stages of the loading represented by dots on the curve.
 After an elastic phase, damage starts to develop at the initial crack tip when $Y$ reach $\Yc$. The reaction force on the top edge of the square then rapidly drops as two cracks develop in a symmetric pattern starting from the crack tip until the reaction reaches zero when the two cracks reach the right boundary. Notice how cracks gradually turn while  advancing until they reach the right boundary. This result could not have been obtained with Griffith theory alone: one should have had some model to predict the crack advance direction. As in the phase field or in the TLS approach, no additional assumption is needed to account for crack direction change. It is the minimization of the energy that naturally drives the crack. This example also shows that the Lip-field approach is able to represent branching cracks.
 The crack pattern and the load/displacement curve is consistent with published results \cite{wu2020phase}, \cite{allaire2011damage}. These results are however unrealistic considering that the top crack develops in compression, so that when $d$ reaches $1$, we have overlapping of the lips of the crack. This issue is not new. In order to avoid this behavior, we need an asymmetric traction/compression model, where the damage function $g(d)$ only affects the traction part of the strain energy as discussed in part \ref{sec:mecha}. The simulations have been re-run using this  change in  and on a mesh adapted for the expected crack path. The results are reported on figure \ref{fig:sheartestld_assym}. 

Up to the point where the damage starts to localize at the initial crack tip, the results are similar to the symmetric case. Then, when the reaction force starts to decrease, the damage zone localizes in only one crack, on the bottom side of the square, propagating first along a straight line until it gets close to the bottom fixed boundary and the reaction force reaches a local minimum. Then it starts to turn while the reaction start growing again, in contrast with the previous case. Notice how the gradient of the damage in the normal direction to the boundary ($\mathbf \grad d \cdot \mathbf n $) is not zero when the crack reaches the boundary. This is in sharp contrast with results obtained in the framework of the phase field theory where the strong form imposes $\mathbf \grad d \cdot \mathbf n  = 0$.


\subsection{Two edge cracks}
\begin{figure}[ht]
	\begin{center}
		\def\svgwidth{0.5\textwidth}
		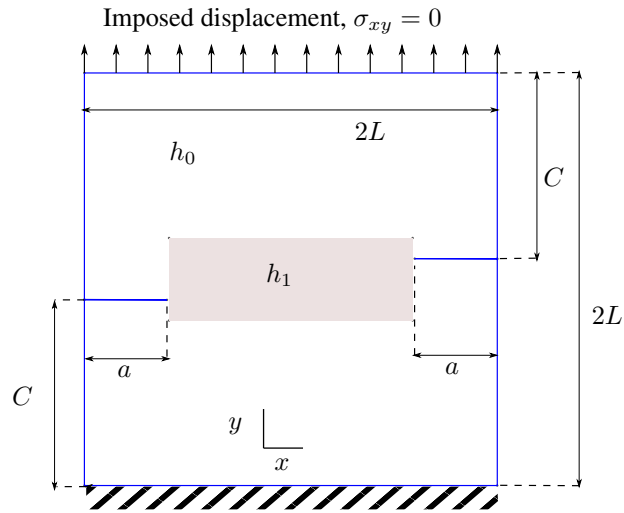
	\end{center}
	\caption{Two edge cracks, geometric description, boundary condition and meshing parameters.}
	\label{fig:twoedgecrack_geom}
\end{figure}

For the last example, we  reproduce another popular test case, two edge cracks, presented and solved for example in \cite{judt2015crack}. Starting from a square sample, two horizontal initial cracks are cut out of the sample, one on each side but at different heights (see figure \ref{fig:twoedgecrack_geom}). 

Table \ref{tab:twoedgecrack_geompar} reports the geometrical, material and mesh parameters used for the simulation. The damage model is here symmetric with regards to traction and compression. As in the previous case, we use two parameters $h_0$ and $h_1$ to control the mesh density. We have a finer mesh in the zone where the damage is expected to propagate (light gray on figure \ref{fig:twoedgecrack_geom}).
\begin{table}[ht]
	\begin{center}
		\begin{tabular}{ll}
		Geometrical parameters: & Mesh parameters: \\
		
		\begin{tabular}{|r|r|r|r|}
			\hline
			  &$L$ & $C$  & $a$\\
			\hline
			mm & 1 & 0.95& 0.2 \\
			\hline  
		\end{tabular} &

		\begin{tabular}{|l|l|l|l|}
			\hline
			 $h_0$ & $h_1$  & vertices & faces\\
			\hline
			  0.025mm & 0.003125mm & 26338 &52675 \\
			\hline  
		\end{tabular}
	\\
	& \\
	&
	Material parameters: \\
	&	
	\begin{tabular}{|l|l|l|l|l|}
		\hline
		$E$ & $\nu $ & $l$ & $\Yc$ & $\eta$ \\
		\hline
		1 MPa &	0.2 & 0.01 mm & 1 MPa & 0.1\\
		\hline  
	\end{tabular}

		\end{tabular}

	\end{center}
	\caption{Two edge cracks geometrical,  mesh and  material parameters.}
	\label{tab:twoedgecrack_geompar}
\end{table}

\begin{figure}[ht]
  \begin{center}
	\includegraphics[width=1.\textwidth]{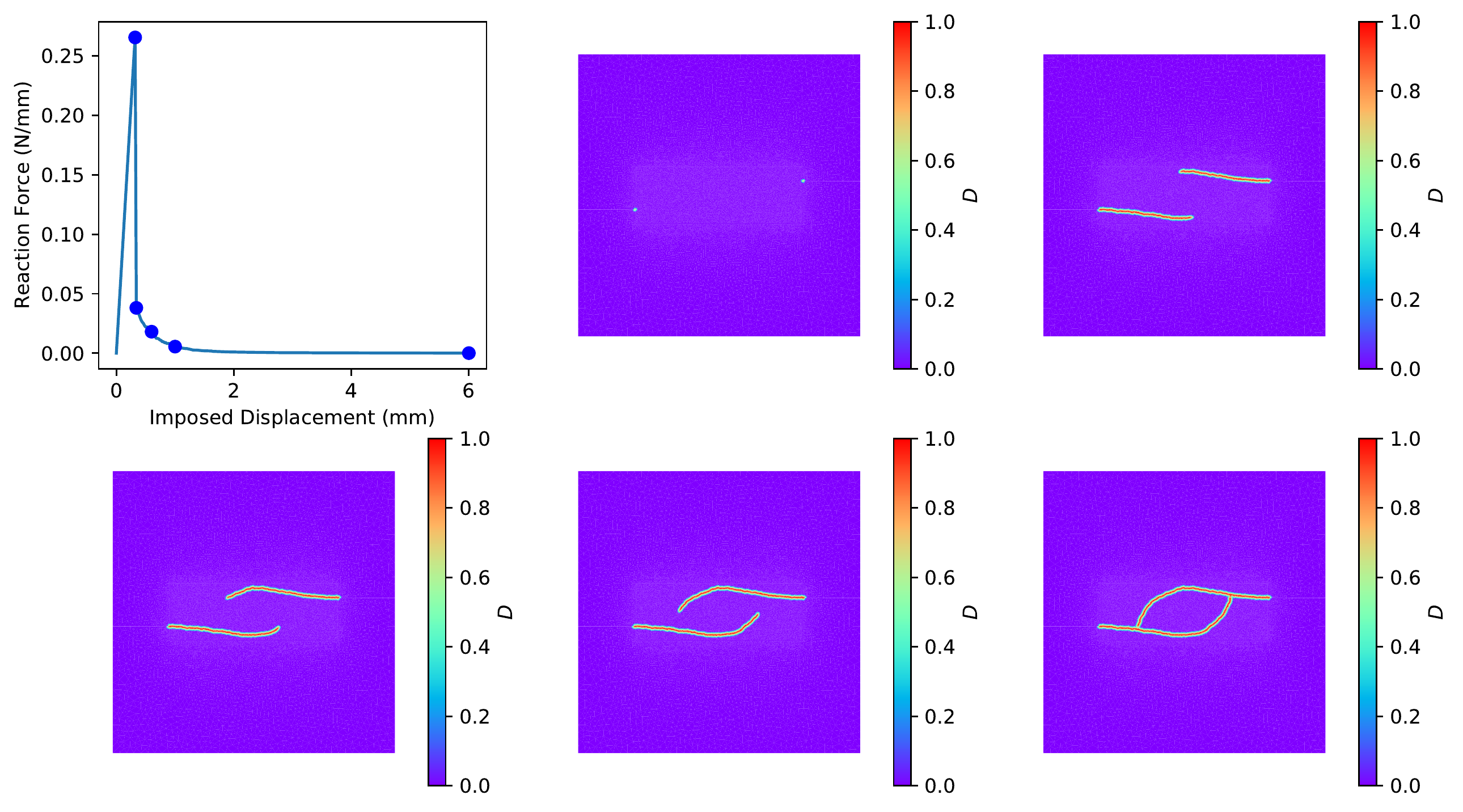}
  \end{center}
\caption{Two edge cracks,  top left: load displacement curve. From left to right, damage field at the corresponding dot on the curve.}
\label{fig:twoedgecrack}
\end{figure}

The square is fixed at the bottom while the top edge is pulled in displacement control in the $y$ direction. Results are reported on figure \ref{fig:twoedgecrack}.

We first have a quasi elastic phase where the damage is confined to the two initial crack tips until we reach a critical load. Beyond the critical load, the two damaged zones start to grow, localizing into two cracks that start to propagate, forming a pattern that maintains a central symmetry with regard to the center of the square, while the reaction force on the top boundary quickly drops. The left crack propagates to the right with a slight angle toward the bottom while the right crack propagates to the left with a slight angle toward the top, until both reaches the vertical axis of symmetry of the square. After that, the path of both crack starts to curve toward the horizontal axis of symmetry of the square until both crack tips finally merge. Note that the damage at a distance larger that $l$ from the place where $d$ reaches its maximum is exactly zero. There is no damage away from the localization zone. The crack pattern obtained is similar to what is observed in previous work.

\section{Conclusions and future work}


The paper described a first two dimensional finite element implementation of the Lip-field approach for brittle fracture with symmetric and asymmetric damage models in tension/compression.
This follows introduction of the Lip-field approach and its  one dimensional implementation reported in \cite{Moes2021lipschitz}. 

A variety of examples have been treated demonstrating the independence of the crack path from the mesh, and good convergence properties.
A comparison with the Griffith model has also been provided and shows good
correlation.

The main originality of the Lip-field approach is to be found in the way the local equations are regularized. Instead of adding a 
damage gradient contribution to the incremental potential, 
as in the phase-field approach, the damage is 
constrained to be Lipschitz continuous 
under a given length. 
This approach has some advantages compared to phase-field. 
The main one is that we can use a local minimization as a starting point. 
Then, using the bounding technique, construct patches  where the Lipschitz constraint needs to be enforced.
This strongly reduces the number of elements where we must explicitly minimize under non local constraint the potential with respect to the $d$ field in the staggered scheme. 
This naturally leads to an algorithm where the cost of the minimization on $d$ can be much lower than the computation of the equilibrium at fixed $d$.

Encouraged by these results, we plan to extend the method in the following directions in the future:
\begin{itemize}
	\item Softening plasticity. This would open a large array of potential applications. We previously demonstrated in the 1D case that the approach was sound and gave interesting results. Extending this approach to two dimensions should not be difficult and should be a low hanging fruit for the method.
	\item Mesh refinement. It is clear from the examples, that as well as phase field, quite fine meshes are necessary to capture the shape of the damaged zone. A logical step toward better results would be to allow automatic mesh refinement to capture the localized damage zone at low cost during a simulation. Strategies to transfer the $d$ field from one mesh to another, while fulfilling the Lipschitz constraints need to be developed.
	\item Improving the resolution step for the $d$ field. Even if the non-local part of the constrained minimization can be done on relatively small patches of the mesh, the resolution algorithm that we used so far might not be optimal. We indeed rely on interior point method  which has the inconvenient to prevent easy use of good starting point for the minimization, because of the centering step typical of such family of methods. Considering that we could obtain good starting points, iterative projection methods might be a better choice for our specific problem. Taking more advantage of the fact that the objective function is separable in $d$ should also help finding a faster algorithm.
	\item Dynamics. It would be interesting to apply the method in a dynamical setting. That would be the occasion to study cases where the crack path become much more complicated.
	\item Finally, we could of course move to three-dimensional problem. The only difficulties are technical: the problems to solve will become big very quickly, and parallel implementation will be needed. A 3 dimensional implementation would need to take advantage of all the improvements cited previously in order to give results in a reasonable computational time.
\end{itemize}

\appendix
\section*{Appendices}
\section{Proof of  \eqref{eq:inclusions} recalled below}\label{appendix:incl}

\begin{equation}
	\Liph  \subset \Liphu \subset \Liphd = \Liphp
\end{equation}
To prove the first inclusion, consider the shortest path inside $\Delta^h$ linking two vertices. 
This path is given by a continuous curve $\bfz(s), 
s \in [0,1]$. Suppose now $d \in \Liph$, we thus have 
\begin{equation}
	-\frac{1}{l} \norm{\frac{\dint \bfz}{\dint s} } \leq 
	\nabla d \cdot \frac{\dint \bfz}{\dint s}   \leq 
	\frac{1}{l} \norm{\frac{\dint \bfz}{\dint s} }
\end{equation}
Integrating the above with respect to $s\in [0,1]$ gives the result. 
The second inclusion in \eqref{eq:inclusions} is a direct consequence of  
the fact that 
\begin{equation}
	\disth(\bfx, \bfy) \leq \disthp(\bfx, \bfy), \quad \forall 
	\bfx, \bfy \in V
	\label{eq:ineqd}
\end{equation}
Finally, regarding the equality in \eqref{eq:inclusions}, we first note that $\Liphd \subset \Liphp$ since $\Liphd$ checks all pairs of vertices whereas $\Liphp$ only checks pairs of vertices 
forming an edge. And for two vertices $\bfx$, $\bfy$ forming an edge, we have 
$\disthp(\bfx, \bfy) = \norm{\bfx - \bfy}$. Then, we prove $\Liphp \subset \Liphd$ using similar arguments as the one used to prove the first inclusion.

\section{An example of  Lipschitz projection}\label{appendix:exampleproj}
As an example we consider a damage field $d$ that does not satisfy the Lipschitz constraint of length scale $l$,
and we wish to project onto $\Lip$ while minimizing the  $\Ltwo $ distance to $d$.
 We define the distance between two field as the usual distance in $\Ltwo$ as  $w(d_1, d_2) = (\int_{\Omega} (d_1 - d_2)^2  d{\Omega})^{1/2}$. 
 The projection of $d$ onto $\Lip$ is defined by $\pi_\Lip(d) =\arg \min_{d' \in \Lip} w(d,d')$.
 
 Let $\Omega$ be a square of length $L$, centered at point at the origin of the 2D plane. We define a field $ d \in \Ltwo $ which is not in $\Lip$ as:
 \begin{equation}
 	d(x,y) = \max \left(1 - r(x,y)/\bar l ,0 \right)
 \end{equation}
 where $r$ is the distance to the origin and $\bar l< l$ is  a constant actually corresponding to the Lipschitz constant of the field $d$. In this case we can easily compute the projection on $\Lip$ as:
 \begin{equation}
   \pi_\Lip(d) = \max \left( \sqrt[3]{ \left(\frac {\bar l}{l}\right)^2 } - r(x,y)/l, 0 \right)
 \end{equation}
 
 Now we want to check the precision of the discretized projection on different mesh, using either the $\Liph$ or $\Liphp$ constraint. For the computation, we use $l=1$ and $\bar l = 1/4$ and a series of structured meshes parametrized by $L/h$, where $h$ is the length of the edges along the $x$ axis.  
 
 \begin{figure}[ht]
 	\begin{center}
 		\includegraphics[width=\textwidth]{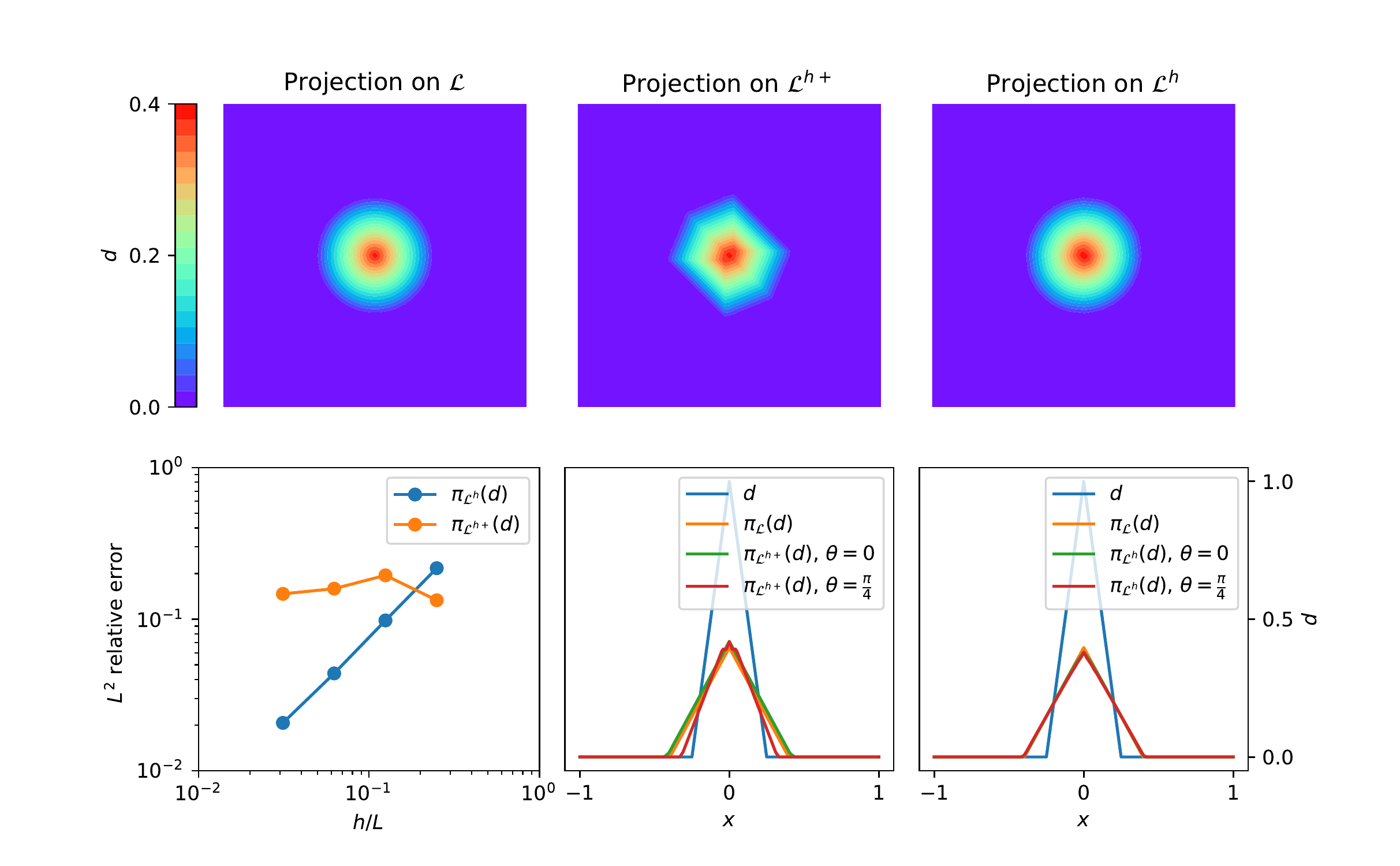}
   \end{center}
 	\caption{Convergence of the discrete lip projection.
 		Top, from left to right: $\pi_\Lip(d)$ interpolated on the mesh, 
 		$\pi_{\Liphp}(d)$ and $\pi_{\Liphp}(d)$. Bottom: from left to right, convergence analysis,  cut of $\pi_{\Liphp}(d)$ and $\pi_{\Liph}(d)$  along the horizontal line ($\theta =0 $) and the diagonal ($\theta = \frac{\pi}{4}$), compared to $d$ and $\pi_\Lip(d)$ .
     }
 	\label{fig:lipproj}
 \end{figure}
 
 Results are reported on figure \ref{fig:lipproj}. On the top row, the map of $\pi_\Lip(d)$, $\pi_{\Liphp}(d)$ and $\pi_{\Liph}(d)$ are plotted for $h/L = 1/32$. 
 On the bottom row, starting from the left, we plotted first the $\Ltwo$ relative error norm for  $\pi_{\Liphp}(d)$ and $\pi_{\Liph}(d)$ as a function of $h/L$, followed by extraction along lines passing trough the origin at different angle $\theta$ from the $x$ direction the values of $d$, and its different projections computed on mesh $h/L = 1/32 $.
 
 The map of $\pi_{\Liph}(d)$ is very close to the map of $\pi_\Lip(d)$, This is confirmed by the cut plot, and the convergence analysis where the error scale linearly with $h$.
 On the contrary, refining the mesh for $\pi_{\Liphp}(d)$ does not improve significantly the results. Indeed, if we get a correct estimate of the maximum value, the slope of the projection slightly change depending on the direction, in a manner strongly dependent with the alignment of the direction with the existing edges in the mesh.

\section{Dijkstra based algorithm to compute the bounds \eqref{eq:bl} and \eqref{eq:bu}} \label{appendix:dijkstra}
The set $V$ of vertices in the $\Delta^h$ mesh is partitioned into 
a set of trial and final vertices, denoted $\Vt$ and $\Vf$, respectively.
At any step in the algorithm, we have $\Vt \cup \Vf = V$ and 
$\Vt \cap \Vf = \emptyset$. The algorithm starts with $\Vf$ empty and ends when $\Vt$ is empty. 
The bounds are initialize to $\dloc$ given by \eqref{eq:dloc}.
The upper bound $\du$ is obtained by
\begin{itemize}
	\item Step 0:  $\Vt = V$, $\Vf = \emptyset$, $\forall v \in V: 
	\du(v)  \leftarrow \dloc(v) $
	\item Step 1: $v^* = \arg \max_{\substack{v \in \Vt}} \du(v)$, 
	$\Vt \leftarrow \Vt \setminus \{ v^* \}$, $\Vf \leftarrow \Vf \cup \{ v^* \}$ 
	\item Step 2: $\forall v \in \Vt$ such that $(v, v^*) \in E: \du(v) \leftarrow \max (\du(v), \du(v) - \norm{v - v^*}/l)$
	\item Step 3: If $\Vt = \emptyset$ end, else go to Step 1.
\end{itemize}
Similarly, one may build the lower bound $\dl$ from the following:
\begin{itemize}
	\item Step 0:  $\Vt = V$, $\Vf = \emptyset$, 
	$\forall v \in V: \dl(v) =\dloc(v), $
	\item Step 1: $v^* = \arg \min_{\substack{v \in \Vt}} \dl(v)$, 
	$\Vt \leftarrow \Vt \setminus \{ v^* \}$, $\Vf \leftarrow \Vf \cup \{ v^* \}$ 
	\item Step 2: $\forall v \in \Vt$ such that $(v, v^*) \in E: \dl(v) \leftarrow \min (\dl(v), \dl(v) + \norm{v - v^*}/l)$
	\item Step 3: If $\Vt = \emptyset$ end, else go to Step 1.
\end{itemize}
To unsure $O(nlog(n))$ efficiency, where $n$ is the size of $V$, the set $\Vt$ is maintained as a descending sorted list according to the value of $\dl(v)$ (resp. $\du$). Step 1 is reduced to take the first (resp. the last) of the list, and at step 2, all the $v$ for which $\dl(v)$ (resp. $\du(v)$) have been updated must be relocated in the list so that the ordering is maintained.

\section{Proof of the bounds in the discrete setting}\label{appendix:proof_bounds}
{eq:disbound}
We prove \eqref{eq:disbound} recalled below
\begin{equation}
	d_n(\bfx) \leq \dl(\bfx) \leq \dloc(\bfx) \leq \du(\bfx) \leq 1, \quad 
	\dl(\bfx) \leq d^+(\bfx) \leq \du(\bfx), \quad \forall \bfx \in V \label{eq:disbound2}
\end{equation}
The first relation is a direct consequence of the initialization of $\dl$ and $\du$ at $\dloc$ and the use of  Step 2 of the algorithm (detailed in the previous appendix).
The second relation in \eqref{eq:disbound2} amounts to prove that the optimal solution cannot be outside the bounds. The same argument as in in the continuum setting \cite{Moes2021lipschitz} may be used. 
The objective function is a sum of convex functions 
in $d$ at every vertex of the $\Delta^h$ mesh whose 
minimum is $\dloc$. 
If the optimal solution lies outside the bound, we can build a better solution by projecting it on the bounds 
(the minimum or maximum of Lipschitz functions stays Lipschitz). This proves the second part of \eqref{eq:disbound2}.

\bibliographystyle{abbrv}
\bibliography{refs.bib} 

\end{document}